\documentstyle{BSAXwork}

\input epsf.sty

\def \AAP #1 #2 {{\em Astron. Astrophys.\/} {\bf #1}, #2}
\def \AAL #1 #2 {{\em Astron. Astrophys. Lett.\/} {\bf #1}, L#2}
\def \AAR #1 #2 {{\em Astron. Astrophys. Rev.\/} {\bf #1}, #2}
\def \AAS #1 #2 {{\em Astron. Astrophys. Suppl. Ser.\/} {\bf #1}, #2}
\def \AJ #1 #2 {{\em Astron. J.\/} {\bf #1}, #2}
\def \ANNREV #1 #2 {{\em Ann. Rev. Astron. Astrophys.\/} {\bf #1}, #2}
\def \APJ #1 #2 {{\em Astrophys. J.\/} {\bf #1}, #2}
\def \APJL #1 #2 {{\em Astrophys. J. Lett.\/} {\bf #1}, L#2}
\def \APJS #1 #2 {{\em Astrophys. J. Suppl.\/} {\bf #1}, #2}
\def \APSS #1 #2 {{\em Astrophys. Space Sci.\/} {\bf #1}, #2}
\def \ASR #1 #2 {{\em Adv. Space Res.\/} {\bf #1}, #2}
\def \BAIC #1 #2 {{\em Bull. Astron. Inst. Czechosl.\/} {\bf #1}, #2}
\def \JSQRT #1 #2 {{\em J. Quant. Spectrosc. Radiat. Transfer\/} {\bf #1}, #2}
\def \MN #1 #2 {{\em Mon. Not. R. Astr. Soc.\/} {\bf #1}, #2}
\def \MEM #1 #2 {{\em Mem. R. Astr. Soc.\/} {\bf #1}, #2}
\def \PLR #1 #2 {{\em Phys. Lett. Rev.\/} {\bf #1}, #2}
\def \PASJ #1 #2 {{\em Publ. Astron. Soc. Japan\/} {\bf #1}, #2}
\def \PASP #1 #2 {{\em Publ. Astr. Soc. Pacific\/} {\bf #1}, #2}
\def \NAT #1 #2 {{\em Nature\/} {\bf #1}, #2}
\def \SAIT #1 #2 {{\em Mem.\ Soc.\ Astron.\ It.\/} {\bf #1}, #2}
\def \MESS #1 #2 {{\em The Messenger\/} {\bf #1}, #2}
\def \ASTRNACH #1 #2 {{\em Astron. Nach.\/} {\bf #1}, #2}
\def\lesssim{\lower4pt\hbox{${\buildrel < \over \sim}$}}
\def\gtrsim{\lower4pt\hbox{${\buildrel > \over \sim}$}}

\begin{opening}

\title{Broadband Spectra and Variability of BL Lacertae in 2000}
\author{M. B\"ottcher$^{1}$\footnote{Chandra Fellow}, 
H. D. Aller$^2$, 
M. F. Aller,$^2$,
O. Mang,$^3$,
C. M. Raiteri$^4$,
M. Ravasio$^5$,
G. Tagliaferri$^5$,
H. Ter\"asranta$^6$,
M. Villata$^4$}
\institute{$^1$Department of Physics and Astronomy, Rice University, Houston,
TX, USA\\
$^2$Astronomy Department, University of Michigan, Ann Arbor, MI, USA\\
$^3$Institut f\"ur Experimentelle und Angewandte Physik, Universit\"at Kiel,
Germany\\
$^4$Osservatorio Astronomico di Torino, Pino Torinese, Italy\\
$^5$Osservatorio Astronomico di Brera, Milano, Italy\\
$^6$Mets\"ahovi Radio Observatory, Helsinki University of Technology, 
Kylm\"al\"a, Finland}
\date{} 
\end{opening}

\begin{document}

\oddpagefooter{}{}{} 
\evenpagefooter{}{}{} 
\medskip  

\begin{abstract} 
We have organized an extensive multiwavelength campaign on 
BL~Lacertae in the second half of 2000. Simultaneous or 
quasi-simultaneous observations were taken at radio frequencies,
in the optical --- carried out by the WEBT collaboration ---,
in X-rays --- using BeppoSAX and RXTE ---, and at VHE gamma-rays 
with the CAT and HEGRA Cherenkov telescope facilities. In this 
paper, we are presenting first results from this campaign.
The WEBT optical campaign achieved an unprecedented time 
coverage, virtually continuous over several 10 -- 20~hour
segments, and revealed intraday variability on time 
scales of $\sim 1.5$~hours. The X-ray observations of Nov.
1 -- 2, 2000, revealed significant variability on similar time 
scales, and provided evidence for the synchrotron spectrum 
extending out to $\sim 10$~keV during that time. At higher
energies, the onset of the hard power-law commonly observed 
in radio-loud quasars and radio-selected BL Lac objects is 
seen. From the energy-resolved X-ray variability patterns,
we find evidence that electron cooling might not be dominated
by one single radiation mechanism; rather, a combination of
synchrotron and Compton cooling seems to be required.
\end{abstract}

\medskip

\section{Introduction}

BL Lacertae, historically the protypical BL~Lac object, has been
the target of many radio, optical, X-ray, and $\gamma$-ray
observations in the past, and has been studied intensively during
various intensive multiwavelength campaigns (e.g., Sambruna et
al. 1999, Bloom et al. 1997, Madejski et al. 1999, Ravasio et al.
2002a). It is a particularly interesting object for detailed 
X-ray studies since it has been repeatedly observed 
that this is where the two broad components of the
multiwavelength SEDs of blazars are overlapping and intersecting:
While the low-frequency component, from radio to soft X-ray frequencies,
is generally believed to be synchrotron emission from relativistic
electrons in a relativistic jet, the high-energy component, at hard
X-rays and $\gamma$-rays, is consistent with Compton emission from
the same population of electrons, scattering their own
synchrotron emission and/or soft photons originating external to
the jet. In agreement with this hypothesis, the X-ray spectrum of
BL~Lacertae has repeatedly shown a concave shape (e.g., Madejski
et al. 1999), with rapid variability, restricted to the low-energy 
excess portion of the spectrum (Ravasio et al. 2002a). 

Whereas BL Lac objects are generally characterized by 
lineless, nonthermal optical spectra, BL~Lacertae has 
occasionally shown strong, broad emission lines (Vermeulen et al.
1995) and is thus no longer considered a typical BL~Lac object.
The existence of emission lines reveals the presence of a
non-negligible amount of scattering material in the broad line
region (BLR), which is consistent with recent spectral
modeling results of Madejski et al. (1999) and B\"ottcher \& Bloom
(2000), requiring a non-negligible external soft photon energy 
density to reproduce the simultaneous broadband spectrum
of BL~Lacertae during its giant outburst in 1997. 

While simultaneous broadband spectra are very useful to constrain
blazar jet models, there still remain severe ambiguities in their
interpretation w.r.t. the dominant electron cooling, injection,
and acceleration mechanisms. The combination of broadband 
spectra with timing and spectral variability information can 
help to break some of these degeneracies. For this reason, we 
have organized an intensive multiwavelength campaign to observe 
BL~Lacertae in the second half of 2000 at as many frequencies 
as possible. We have put special emphasis on good timing 
information, which we have obtained, in particular, at 
optical and X-ray frequencies. In Section \ref{observations}, 
we briefly review the observations carried out during the 
campaign. The diverse variability patterns are presented 
and analyzed in Section \ref{variability}. The most 
detailed simultaneous broadband spectrum obtained during 
the campaign is discussed in Section \ref{spectrum}. 

\section{\label{observations}Observations}

The broadband campaign on BL~Lacertae in 2000 encompassed
observations at radio, optical, X-ray, and VHE $\gamma$-ray
energies during the period July 17, 2000, until the end of
2000. At radio frequencies, the object was monitored
using the University of Michigan Radio Astronomy Observatory
(UMRAO) 26~m telescope, at 4.8, 8, and 14.5~GHz, and with the
14~m Mets\"ahovi Radio Telescope of Helsinki University, at 22 
and 37~GHz. 

\begin{figure}
\epsfysize=6.1cm
\epsffile{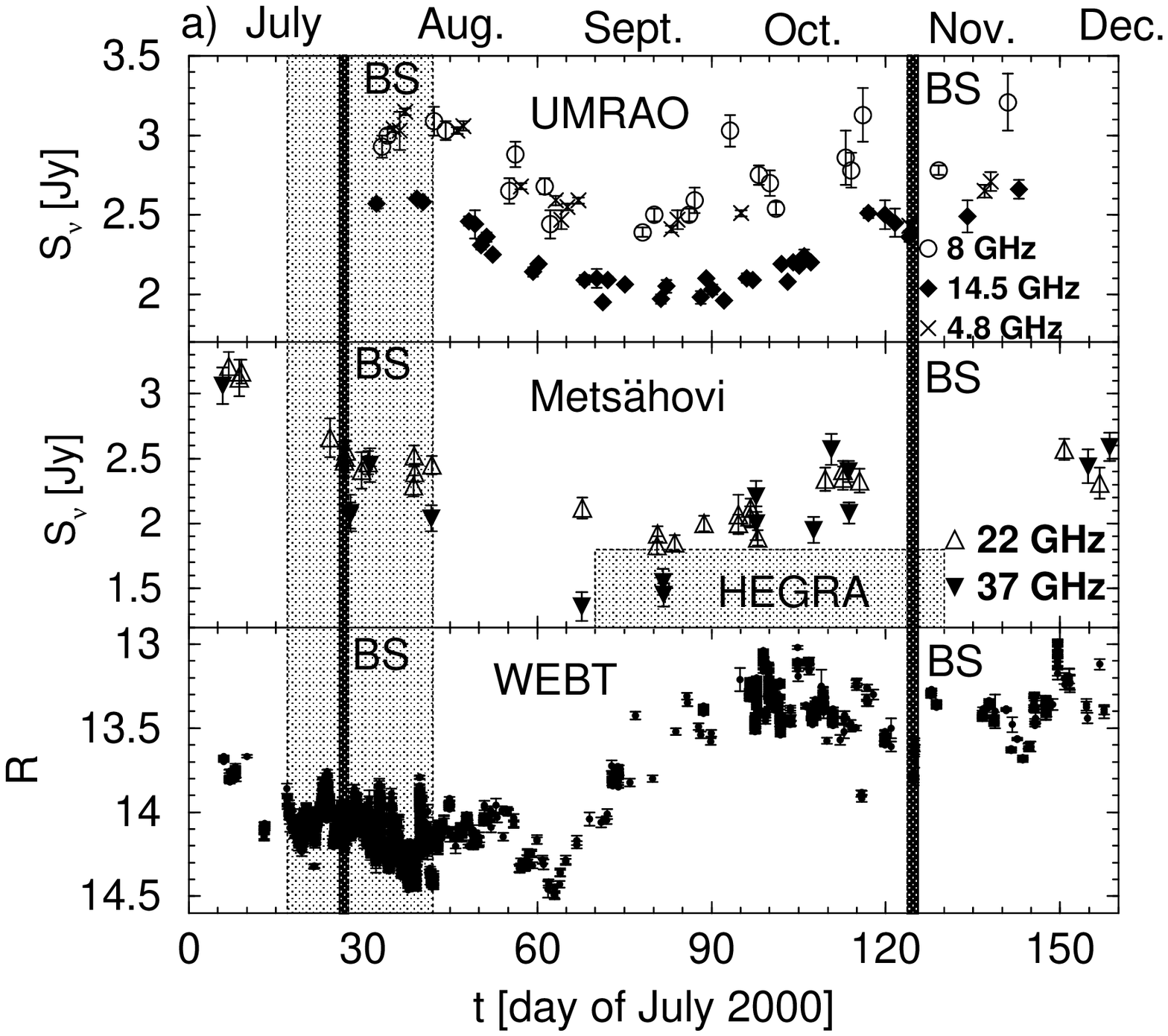}\hskip 0.1cm\epsfysize=5.9cm\epsffile{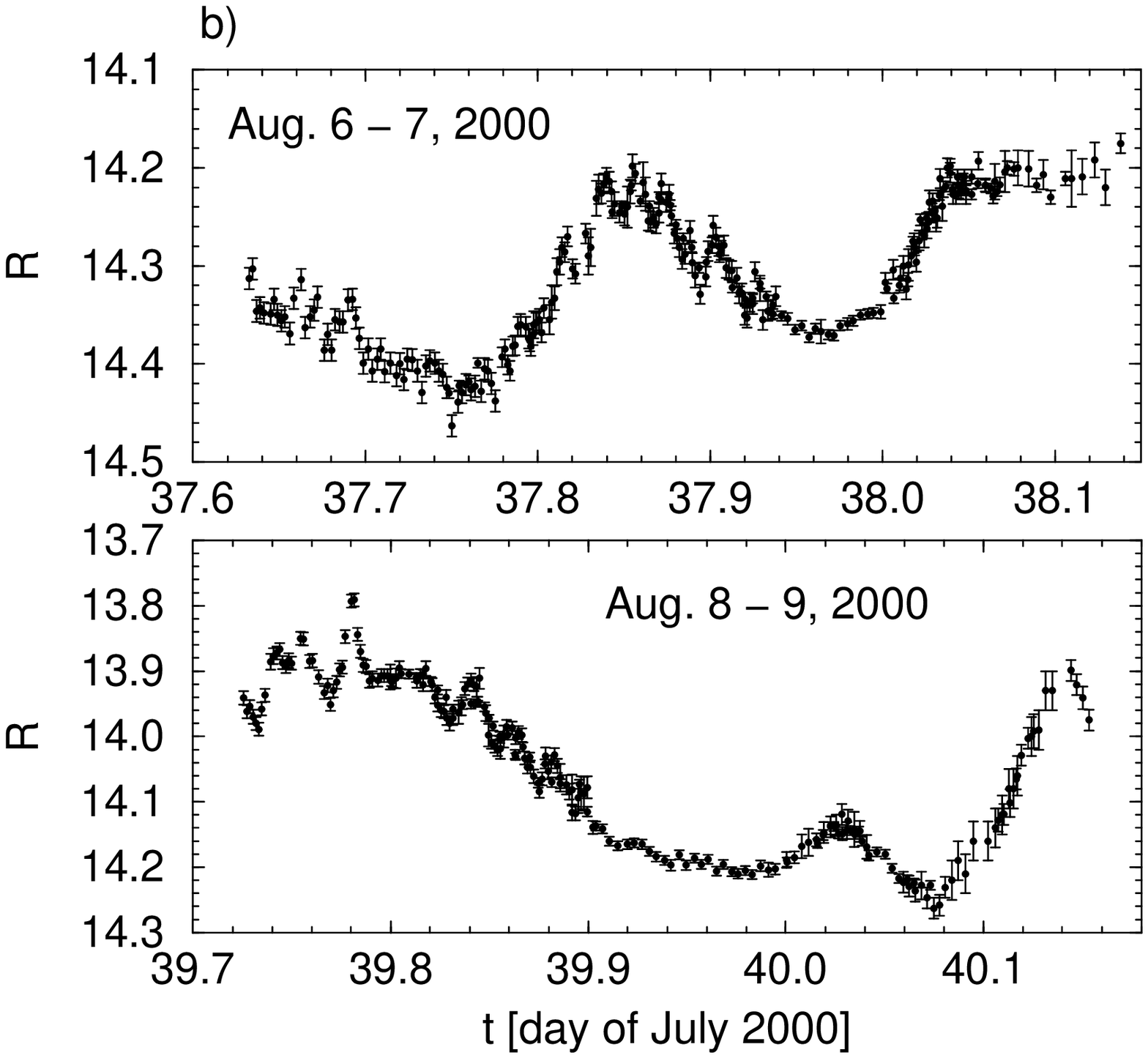}
\caption[]{Left panel (a): Radio and optical R-band light curves 
of BL~Lacertae during the campaign of 2000. The shaded region 
indicates the original core campaign, July 17 -- Aug. 11. 
Also marked are the times of the BeppoSAX pointings (BS) and the 
HEGRA VHE $\gamma$-ray observations.
Right panel (b): Optical (R-band) light curves of BL~Lacertae 
during two nights in August 2000.}
\label{lc}
\end{figure}

Focusing on a core campaign period July 17 -- Aug. 11, 
BL~Lacertae was the target of an intensive 
optical campaign by the Whole Earth Blazar Telescope (WEBT, 
Villata et al. 2000), in which 24 optical telescopes 
throughout the northern hemisphere participated. For more 
details of the WEBT campaign and its results, see 
Villata et al. (2002; see also Raiteri et al. 2002 and 
{\tt http://www.to.astro.it/Groups/Extragal/2200.htm}). 
The radio and optical R-band light curves, along with the
overall timeline of the campaign, are shown in Fig. \ref{lc}a.
The figure illustrates that BL~Lacertae underwent a transition
from a rather quiescent state in July and August 2000 to an
extended high state in mid-September 2000. For this reason, 
the WEBT campaign was extended until the end of 2000, 
although with less dense time coverage than during the  
core campaign.

At X-ray energies, BL~Lacertae was observed with the BeppoSAX
Narrow Field Instruments (NFI) in the energy range 0.1 -- 300~keV
in two 25~ksec pointings on July 26 -- 27 and Nov. 1 -- 2, 2000
(Ravasio et al. 2002b). In addition, the source was monitored 
by the Rossi X-ray Timing Explorer (RXTE) in 3 short pointings 
per week (Marscher et al. 2002); however, the results of those 
observations were not yet available at the time of this 
workshop, and will be included in the final analysis of 
this campaign later. 

BL~Lacertae has been observed by the HEGRA system of imaging
Cherenkov telescopes, accumulating a total of 10.5~h of on-source
time in Sept. -- Nov. 2000. The source was not detected above a
99~\% CL upper limit of 25~\% of the Crab flux at photon 
energies above 0.7~TeV (Mang et al. 2001). The Cherenkov 
Array at Th\'emis (CAT) had also observed BL~Lacertae during 4 
nights between July 26 and Aug. 8 for a total of $\sim 2$~hr. 
However, due to poor weather conditions, those 
observations did not yield useful flux constraints. We
had also been awarded observations with OSSE and COMPTEL on
board the {\it Compton Gamma-Ray Observatory} (CGRO); however, 
unfortunately, the CGRO was de-orbited just around the
time of our scheduled observations.

\section{\label{variability}Radio, Optical and X-ray Variability}

The radio and optical light curves of BL~Lacertae over the 
entire campaign period are shown in Fig. \ref{lc}a. We have
calculated the discrete cross-correlation functions between the
optical and radio bands as well as between the light curves at
different radio frequencies. We found delays between
different radio frequencies with the lower-frequency variability
lagging behind the higher-frequency light curves by up to $\sim
25$~days, with the lags generally increasing with increasing
frequency separation. The frequency dependence of the lags is 
not well described by power-law fits, but rather shows a 
gradual flattening towards the lowest frequencies. 
We found evidence for a positive correlation between the optical 
and radio light curves, indicating a delay of the radio light 
curves with respect to the R band of $\sim 45$ -- 50~d. However, 
our data are spanning only a few months. Bregman et al. (1990) 
had found typical radio -- optical correlations in BL~Lacertae 
with delays of $\sim 1$ -- 4.5~years, 
so that we will have to combine our intermediate-term light curves 
with long-term, historical light curves before drawing firm
conclusions from these delays.

The WEBT campaign returned optical (R-band) light curves of
unprecedented time coverage and resolution. Two examples of
the observed rapid microvariability are shown in Fig. \ref{lc}b.
Brightness variations of $\Delta R \sim 0.35$, corresponding to
flux variations of $(\Delta F)/F \sim 0.4$, within $\sim 1.5$~hr
have been found, which is not exceptional for this source, and
might indicate that BL~Lacertae was in a rather quiescent state. 
However, this does place a constraint on the size of the emitting 
region of $R \lesssim 1.6 \times 10^{14} \, D$~cm, where 
$D = \left(\Gamma [ 1 - \beta\cos\theta_{\rm obs}] \right)^{-1}$
is the Doppler beaming factor. 

\begin{figure}
\epsfysize=5.5cm
\epsffile{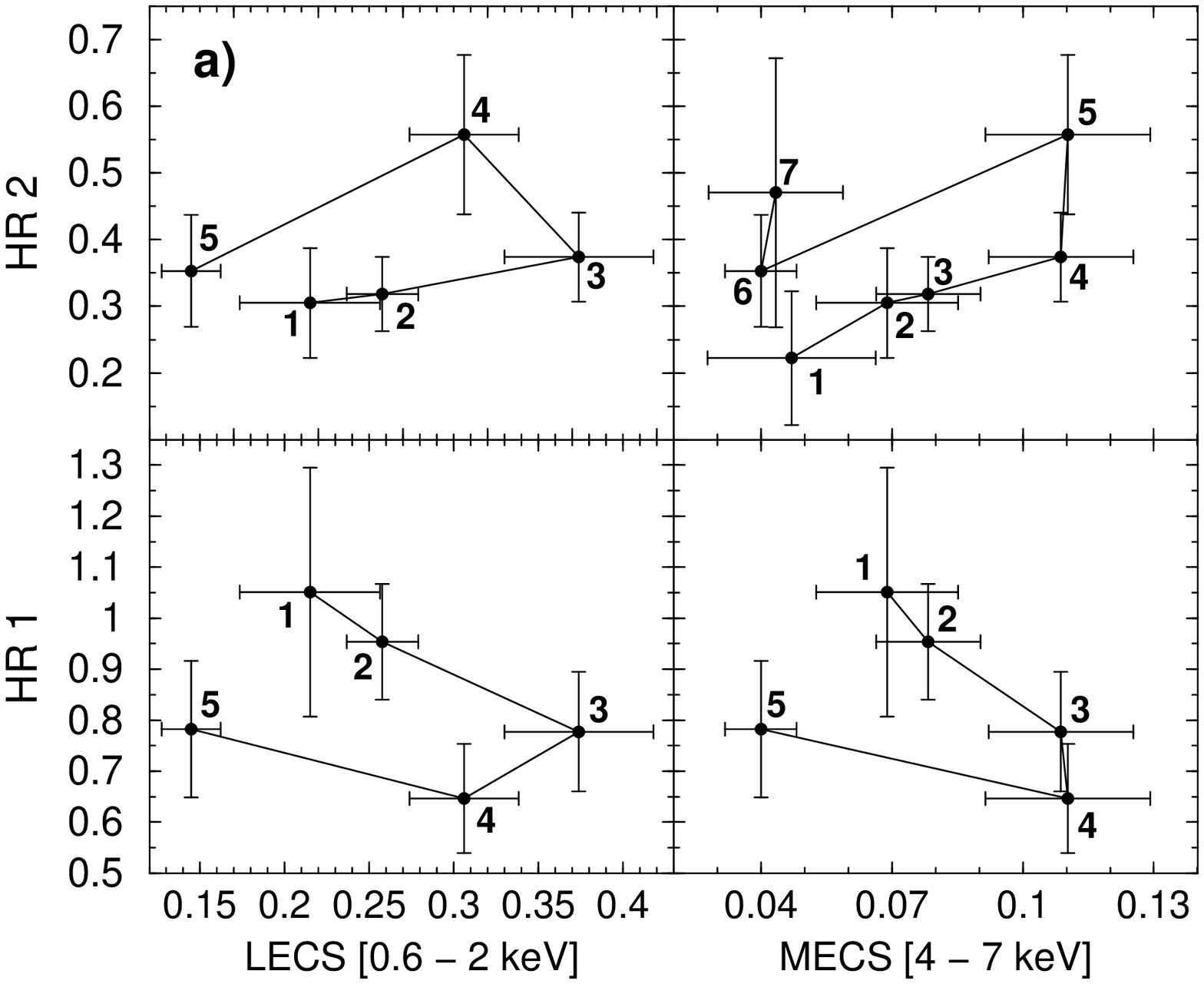}\hskip 0.4cm\epsfysize=5.3cm\epsffile{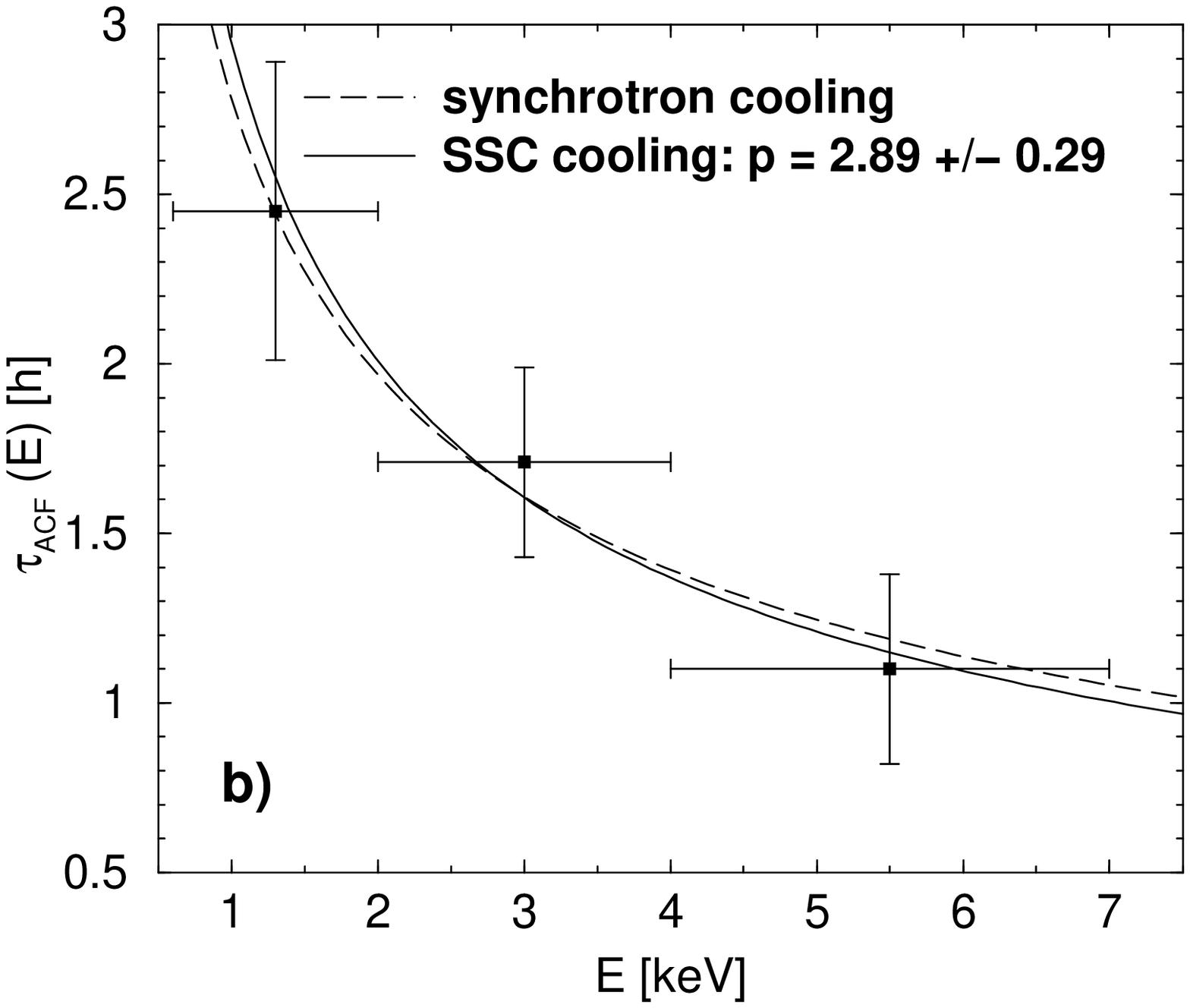}
\caption[]{Left panel (a): Example of the hardness -- intensity 
correlations between the LECS and the hard MECS fluxes and the 
hardness ratios HR1 $\equiv$ (LECS [2 - 4 keV])/(LECS [0.6 - 2 keV]) 
and HR2 $\equiv$ (MECS [4 - 7 keV])/(MECS [2 - 4 keV]). The data
span a maximum 3.5~h from the first to the last point, starting
at 22:30 UT on Nov. 1, 2000.
Right panel (b): Best fits of the cooling characteristic functions
(Eq. \ref{cooling}) to the measured energy-dependent ACF widths.}
\label{xrayvar}
\end{figure}

From the {\it BeppoSAX} X-ray data, we have extracted light curves
with 10-minute sampling, in the energy bands (0.6 -- 2)~keV (LECS),
(2 -- 4)~keV (MECS), and (4 -- 7)~keV (MECS). We also calculated
the hardness ratios HR1 $\equiv$ (LECS [2 - 4 keV])/(LECS [0.6 - 2 keV]) 
and HR2 $\equiv$ (MECS [4 - 7 keV])/(MECS [2 - 4 keV]) with the same
10-minute sampling (see Ravasio et al. 2002b). During our first pointing
on July 26 -- 27, no evidence for short-term variability was found.
The spectrum was hard (energy index $\alpha = 0.76 \pm 0.14$), consistent
with the hypothesis that the entire {\it BeppoSAX} spectrum was dominated
by the SSC emission from low-energy electrons in the jet. 

The results of our second {\it BeppoSAX} pointing on Nov. 1 -- 2 were
quite different. The light curves in all three LECS and MECS energy 
bands exhibited significant variability. Flux variations by a factor 
of $\sim 2$ within $\sim 30$~minutes were detected. The LECS + MECS 
spectrum was well fitted with a steep power-law with energy index 
$\alpha = 1.6$ up to $\sim 10$~keV; the PDS spectrum above 
$\sim 20$~keV indicated a significant spectral upturn, consistent 
with the onset of the SSC component around $\sim 20$~keV. The 
following discussion of the X-ray variability and the 
broadband spectrum of BL~Lacertae during our campaign will focus 
on the results of this second observation on Nov. 1 -- 2.

From the measured fluxes and hardness ratios, we have constructed
hardness-intensity correlations (HICs). As an example, we show in
Fig. \ref{xrayvar}a the various HICs for an individual flare, spanning
a total of $\sim 3.5$~hr. Whereas in the HICs at soft X-ray energies, 
there is some indication of the typical clockwise rotation patterns 
which have also been observed in the X-ray spectral variability of 
high-frequency peaked BL~Lac objects (HBLs, e.g., Mrk~421: Takahashi 
et al. 1996), we also find some correlations indicating the opposite 
behavior (i.e. counter-clockwise rotation patterns) as well as
straight hardness-intensity anti-correlations in some of our soft
X-ray HICs. Interestingly, such diverse HIC patterns have been predicted
through numerical simulations of HBL flaring behaviour in a pure SSC 
model by Li \& Kusunose (2000) under certain conditions: As long as
the electron energy losses are dominated by synchrotron cooling, the 
synchrotron part of the spectrum was predicted to exhibit clockwise
HIC rotation patterns; when SSC cooling becomes important, the 
synchrotron HICs turn into straight anti-correlations. 

Our currently available {\it BeppoSAX} data have too 
large error bars due to limited photon statistics to draw firm 
conclusions on the basis of the measured HICs. However, if those
results can be confirmed by future observations with 
{\it Chandra} or {\it XMM-Newton}, they may provide a new
tool to constrain the dominant cooling mechanism of the electrons
responsible for the synchrotron emission of BL~Lacertae at soft
X-ray energies.

Additional constraints on the physical parameters of the 
emitting region and on the dominant electron cooling mechanism
can be obtained if one can determine the photon energy dependence
of the decay time scale of individual flares. In the two extreme
cases in which either synchrotron or SSC emission is dominating
the electron cooling, the cooling time scale $\tau_c$ (in the
observer's frame) of electrons emitting synchrotron radiation
at a photon energy $E_{\rm sy} = 1 \, E_{\rm keV}$~keV, is

\begin{equation}
\tau_c = \cases{\tau_1 (p, \tau_{\rm T}, \gamma_1) \, B_G^{-p/2} \, 
D^{(2 - p)/2} \; E_{\rm keV}^{(p - 4)/2} 
& for SSC cooling \cr\cr
3.22 \times 10^3 \, B_G^{-3/2} \, D^{-1/2} \, E_{\rm keV}^{-1/2} \; {\rm s}
& for sy. cooling \cr}
\label{cooling}
\end{equation}
(Chiang \& B\"ottcher 2002), where $\tau_1$ is a normalization
factor depending on the spectral index $p$ and the low-energy
cutoff $\gamma_1$ of the injected electron distribution as well
as the Thomson depth $\tau_{\rm T}$ of the emitting region. The 
available data do not have sufficient photon statistics to allow
us to exploit these relations for individual flares. However,
an alternative route to extract information about the average
physical conditions responsible for the observed X-ray flares, is
to compute the autocorrelation functions of the X-ray variability
in the three LECS + MECS energy bands and represent the
widths of those autocorrelation functions as $\tau_{\rm ACF} (E)
= \tau_0 + \tau_c (E)$, where $\tau_0$ corresponds to a
minimum width given primarily by the light-travel time through the
source and the electron injection time scale. The result of this
procedure is illustrated in Fig. \ref{xrayvar}b. Obviously, we can 
not distinguish between the two extreme cases of synchrotron vs.
SSC cooling on the basis of the quality of the fit. However,
important insight can be gained by deriving an estimate of the
magnetic field $B$ from the normalizations of the two fits. For
the case of pure synchrotron cooling, we find $B = (51.2 \pm 5.4)
\, (D/10)^{-1/2}$~G, while for dominant SSC cooling, we find
$B = (106 \pm 12) \, (\tau_{\rm T} / 10^{-6})^{-0.69} \, 
(\gamma_1 / 100)^{-1.31} \, (D / 10)^{0.31}$~G. This is at least
an order of magnitude larger than the magnetic-field values found
in earlier analyses of broadband spectra of BL~Lacertae (e.g.,
Madejski et al. 1999, B\"ottcher \& Bloom 2000). 

One can derive an independent estimate of the magnetic field 
based on the optical synchrotron flux at the time of the 
X-ray observations. Assuming that the $\nu F_{\nu}$ peak of the
synchrotron spectrum is in the optical domain during this period
(see section \ref{spectrum}), and that the magnetic field is in
approximate equipartition with the electrons in the jet, we find
$B_{\rm ep} \approx 8.7 (D / 10)^{-1}$~G. Again, this is about an
order of magnitude lower than the two estimates derived from the 
X-ray variability in the extreme cases of pure synchrotron or pure
SSC cooling. This might be another indication that the electron
cooling is probably not dominated by one single mechanism, and that
possibly additional electron cooling due to Compton scattering of
external soft photons could be required in order to reconcile
the variability and spectral properties found in our observations.

\section{\label{spectrum}Broadband spectrum}

\begin{figure}
\epsfysize=7cm
\begin{center}
\epsffile{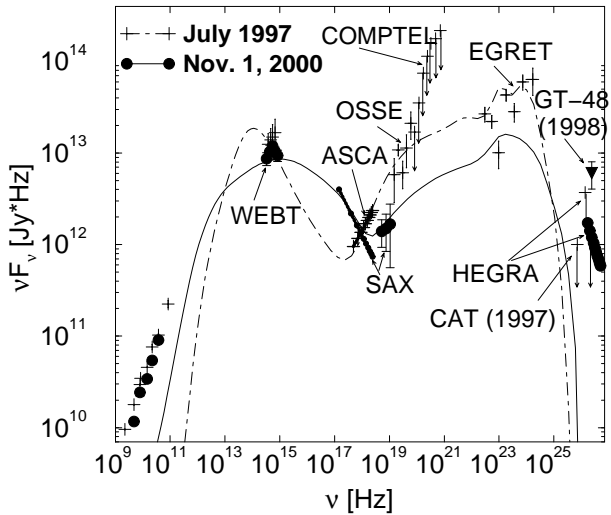}
\end{center}
\caption[]{Simultaneous broadband spectrum of BL~Lacertae from
Nov. 1, 2000 (filled circles), compared to the major outburst
in July 1997 (crosses). Both spectra have been fitted with the
time-dependent jet radiation transfer code described in B\"ottcher 
\& Bloom (2000).}
\label{sp_plot}
\end{figure}

The simultaneous broadband spectrum of BL~Lacertae during the
second BeppoSAX pointing on Nov. 1 -- 2, 2000, is shown in 
Fig. \ref{sp_plot}. For comparison, we also show the spectrum
measured during the source's major outburst in July 1997 (Bloom
et al. 1997). The optical (UBVRI) photometric data have been
de-reddened using the de-reddening law of Cardelli et al. (1989)
with $A_V = 1.2$ (Sambruna et al. 1999) and $R_V = 3.1$. The 
error bars on the optical data points represent the range of 
variability measured during those two days in the respective
optical bands. 

The HEGRA upper limit on the integrated $> 0.7$~TeV photon flux 
was converted to a $\nu F_{\nu}$ flux limit by assuming a power-law
shape of the spectrum with photon index $\Gamma_{\rm ph}$, which 
leads to $\nu F_{\nu} < (\Gamma_{\rm ph} - 1) \times 8.65 \times 
10^{11}$~Jy~Hz at 0.7~TeV. The resulting limiting spectrum is 
shown for $\Gamma_{\rm ph} = 2.5$ in Fig. \ref{sp_plot}. 

The comparison between the two spectra of BL~Lacertae illustrates
the dramatic spectral change at soft to medium-energy X-rays, and 
clearly demonstrates that we are observing the high-energy end of 
the low-frequency (synchrotron) component in the LECS + MECS spectrum.
It is particularly interesting to note that the best-fit spectral 
index of $\alpha = 1.6$ measured at soft X-rays is very close to
the canonical value of $\alpha_{\rm SSC-c}$ expected for the
time-averaged synchrotron spectrum of strongly cooling electrons 
if the energy losses are dominated by SSC cooling (Chiang \&
B\"ottcher 2002). 

The spectrum has been modeled with the time-dependent jet radiation
transfer code described in B\"ottcher \& Bloom (2000). The most
relevant changes of the best-fit model parameters with respect to
the fit to the 1997 flare spectrum correspond to an extension of
the injected electron spectrum towards both lower and higher energies
(i.e. decreasing $\gamma_1$ and increasing $\gamma_2$). Our spectral
modeling results are consistent with the tentative result from our
timing analysis that synchrotron, SSC, and external-Compton cooling
all contribute significantly to the bolometric luminosity of the
source and, consequently, to the electron cooling.

\acknowledgements
The work of M.B. is supported by NASA through Chandra Postdoctoral
Fellowship Award no. 9-10007, issued by the Chandra X-ray Center,
which is operated by the Smithsonian Astrophysical Observatory for
and on behalf of NASA under contract NAS~8-39073.

\end{document}